# An Efficient Soft Decoder of Block Codes Based on Compact Genetic Algorithm

**Ahmed Azouaoui, Ahlam Berkani and Pr. Mostafa Belkasmi**

**SIME Lab., ENSIAS, Mohammed V-Souissi University, Rabat, Morocco**

**Abstract**

Soft-decision decoding is an NP-hard problem with great interest to developers of communication systems. We present an efficient soft-decision decoder of linear block codes based on compact genetic algorithm (cGA) and compare its performances with various other decoding algorithms including Shakeel algorithm. The proposed algorithm uses the dual code in contrast to Shakeel algorithm which uses the code itself. Hence, this new approach reduces the decoding complexity of high rates codes. The complexity and an optimized version of this new algorithm are also presented and discussed.

## 1. Introduction

In digital communication, one of the important issues is how to transmit the message from the source to the destination as faithfully as possible. One of the most used techniques and also the most convenient is the adoption of error-correcting codes. Indeed the codes are used to improve the reliability of data transmission over communication channels susceptible to noise. The coding techniques are based on the following principle: add the redundancy to the transmitted message to obtain a vector called "code word". Decoding techniques are based on the algorithms witch try to find the most likely transmitted code word related to the received one (see Fig. 1).

Decoding algorithms are classified into two categories: hard-decision and soft-decision algorithms. Hard-decision algorithms work on a binary form of the received information. In contrast, soft decision algorithms work directly on the received symbols [1].

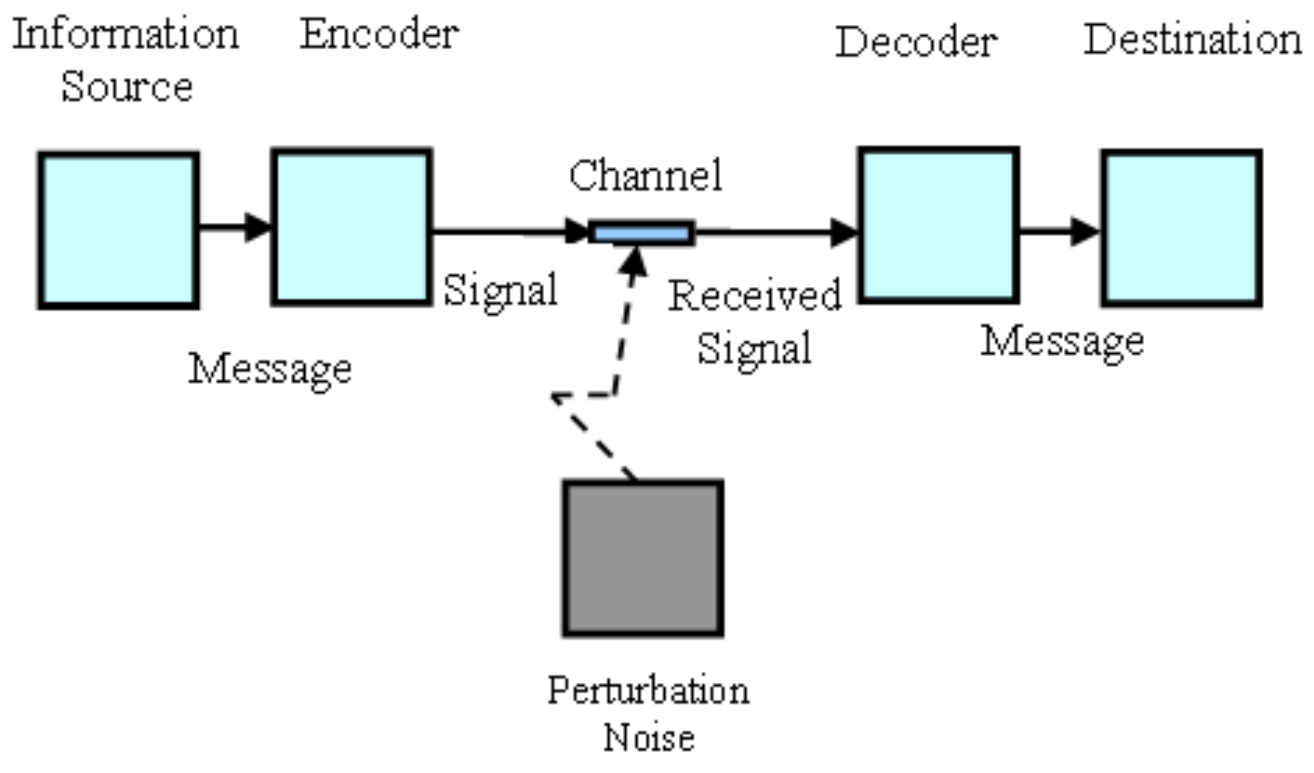

Fig. 1. A simplified communication system model.

Soft-decision decoding is an NP-hard problem and was approached in different ways. Recently artificial intelligence (AI) techniques were introduced to solve this problem. These techniques show very good results. Among related works, one work A* algorithm to decode linear block codes [2], another one uses genetic algorithms (GA) for decoding linear block codes [3] and a third one uses compact genetic algorithms to decode BCH codes[4]. Maini and al. [5] were the first, according to our knowledge, to introduce Genetic algorithms in linear block codes decoding. Hebbes and al. [6] worked on the integration of genetic algorithms in a classical turbo codes decoder, and Durand and al. [7] worked on the optimization of turbo decoding by optimizing the interleaver with a genetic algorithm. Furthermore the deployment of Artificial Neural Networks (ANN), to train the system for higher fault tolerance in OFDM is used by Praveenkumar [8]. There are also other works [9-11] based on AI trying to solve problems related to coding theory. All these decoders based on GA use the generator matrix of the code; this fact makes the decoding very complicated for codes of high rates.

We have investigated the use of genetic algorithms in different ways. In [12], GA is used to search good double-circulant codes. In [13], a new soft decoder of block codes based on the classical genetic algorithm with very good performances was presented.

The Compact Genetic Algorithm Decoder (CGAD) is a significant contribution to soft-decision decoding. In effect a comparison with other decoders, that are currently the most successful algorithms for soft decision decoding, shows its efficiency. This new decoder can be applied to any binary linear block code, particularly for codes without algebraic decoder. Unlike Chase algorithm which needs an algebraic hard-decision decoder. Further, it uses the dual code and work with the parity-check matrix. The later makes them less complicated for codes of high rates. In order to show the effectiveness of this decoder, we applied it for BCH, QR and RS codes over AWGN transmission channel.

The remainder of this paper is organized as follows: in section 2, we introduce the compact genetic algorithm. In section 3, CGAD, our genetic algorithm for decoding, is described. Section 4 reports the simulation results and discussions .In sectionI 5, we study the complexity of our algorithm.The optimized version of the proposed algorithm will be presented in section 6. Finally, Section 7 presents the conclusion.

## 2. The Compact Genetic Algorithm

The Compact Genetic Algorithm (cGA), proposed by Harik and al. [19], is a special class of genetic algorithms. It represents the population as a probability distribution over the set of solutions; thus, the whole population do not need to be stored. At each generation, cGA samples individuals according to the probabilities specified in the probability vector. The individuals are evaluated and the probability vector is updated towards the better individual. Hence, its limitation hinges on the assumption of the independency between each individual bit. The cGA has an advantage of using a small amount of memory. The pseudo code of cGA is shown in Fig. 2. The parameters are the step size $(1/\lambda)$ and the chromosome length (l).

First, the probability vector $p$ is initialized to 0.5. Next, the individuals $a$ and $b$ are generated from $p$. The fitness values are then assigned to $a$ and $b$. The probability vector is updated towards the better individual. In the population of size $\lambda$, the updating step size is

```
1. Initialize the probability vector
 For i := 1 to l do p[i] := 0.5;
2. Generate two individuals from the probability vector
  a := generate(p);
  b := generate(p);
3. Let them compete
 winner, loser := compete(a, b);
4. Update the probability vector towards the better one
 For i := 1 to l do
        if winner[i] ≠ loser[i] then
            if winner[i] = 1 then p[i] := p[i] + 1/λ
                      else  p[i] := p[i] – 1/λ
5. Check if the vector has converged
 For i := 1 to l do
        if p[i] > 0 and p[i] < 1 then
         return to step 2;
```

Fig.2 Pseudo code of cGA

$1/\lambda$ ; The probability vector is increased or decreased by this size. The loop is repeated until the vector convergence.

## 3. The Algorithm Proposed

Consider a linear block code $C$ of length $n$, dimension $k$ and minimum Hamming distance $d$, defined over the Galois field of order 2 (GF (2)). Also, let $H$ be the parity matrix of $C$. It is assumed that code words of $C$ are modulated by a BPSK modulator and transmitted over an AWGN channel. Let $r = (r_1, r_2, \cdots, r_n)$ be the received word from the output of channel. Upon receiving the vector $r$, the demodulator makes hard-decisions $w_i$,

$$i = 1, \cdots, n, \qquad w_i = \begin{cases} 1, & r_i \geq 0, \\ 0, & r_i \prec 0 \end{cases}$$

The receiver then calculates the syndrome $wH$ and accepts $w$ as the most likely transmitted codeword if the syndrome $wH = 0$. If the syndrome $wH \neq 0$, the soft decoding process begins as described below:

**Step 1**: Sorting the sequence $r$ in such a way that $|r_i| > |r_{i+1}|$ for $1 \leq i \leq n$. Further, permute the coordinates of $r$ to ensure that the last $(n-k)$ positions of $r$ are the least reliable linearly independent positions. Call this vector $r'$ and let $\pi$ the permutation related to this vector $\left(r' = \pi(r)\right)$. Apply the permutation $\pi$ to $H$ to get a new check matrix $H' = [A I_{n-k}]\,(H' = \pi(H))$.

**Step 2**: Define the objective function:

An individual is a set of k bits. Let $E'$ be an individual, $z$ be the quantization of $r'$, $S$ be the syndrome of $z$ such that $S = zH'$, $S_1$ be an $(n-k)$ -tuple such that $S_1 = E'A$ where $A$ is sub matrix of $H'$, and $S_2$ be an $(n-k)$ -tuple such that $S_2 = S + S_1$. We form the $E$ error pattern such that $E = (E', E'')$, where $E'$ is the chosen individual and $E'' = S_2$. Then, $z + E$ is a code word.

The fitness function is the correlation discrepancy between the permuted received word and the estimated error such that:

$$f_{r'}(E) = \sum_{j=1, E_j=1}^{n} |r'_j| \quad (1)$$

**Step 3**: Map $r'$ onto probability vector $p$, $p \in R^k$

The probability vector $p$ defines the starting point for the genetic search over the $k$ -dimensional vector space $F_2^k$. It is expected that this search would terminate (converge) at a vertex of the $k$ -dimensional hypercube. An obvious starting point is the center point of the search space, i.e. $p = \{0.5\}_1^k$. However, the search time and complexity can be greatly reduced if the search is initiated from a point close to the solution vector. The following steps describe a method that uses soft information of the received vector to determine a starting point close to the optimum solution.

**Step 4**: Generate a pair of binary random vectors $a_1$, $b_1$.

A pair of vectors $a_1, b_1 \in F_2^k$ is generated with the following probabilities:

$$P(a_i = 1) = p_i,$$
$$P(a_i = 0) = 1 - p_i.$$

These vectors can be generated using an uniform random number generator $U(0,1)$, as follows:

$$a_i = \begin{cases} 1 & U(0,1) \prec p_i \\ 0, & otherwise. \end{cases}$$

**Step 5**: Evaluate $F(a_i)$. This step evaluates the fitness values of $a_1$ and $b_1$ using the objective function (1). The vector with the greater value is identified as the winner $\alpha$ while the vector with the lesser value is identified as the loser $\beta$.

**Step 6**: Update the probability vector $p$.

The vector $p$, is updated towards the fitter one (winner $\alpha$) using the following rules:

- if $\alpha_i = 1$ *and* $\beta_i = 0$, *then* $p_i = p_i + 1/\lambda$
- if $\alpha_i = 0$ *and* $\beta_i = 1$, *then* $p_i = p_i - 1/\lambda$
- if $\alpha_i = \beta_i$ *then* $p_i$ *not will be updated*.

The updating step size $1/\lambda$ is a user defined parameter which is directly related to the performance of the decoder.

**Step 7**: Converged? i.e. $(p_i = 0 \text{ or } p_i = 1)$

If No, go to Step 4.

If yes, the converged $p$ gives the final solution for the objective function (1).

**Step 8**: Encode $p$ and apply inverse permutation.

**Remark:**

In step 1 of the CGAD, in order to have a light algorithm we apply the Gaussian eliminations on the independent columns corresponding to the least reliable positions, without the permutation. This optimization is not used in other similar works [3, 4, 5].

## 4. Performances Study

In order to show the effectiveness of CGAD, we do intensive simulations. For transmission we used an AWGN channel with a BPSK modulation. The simulations where made with default parameters outlined in Table 1.

Table 1: Default parameters.

| Parameter | Value |
|---|---|
| $1/\lambda$ | 1/500 |
| Default code | BCH(63,51,5) |
| Channel | AWGN |
| Modulation | BPSK |
| Minimum number of transmitted blocks | 1000 |
| Minimum number of residual bit errors | 200 |

### 3.1 Effect of Step Size on Performances

We illustrate the relation between bit error probability and the step size of CGAD.
The Fig. 3 emphasizes the influence of the step size on the performances of CGAD. The results show that decreasing $1/\lambda$ also improves BER performances. When decreasing the step size from 1/50 to 1/500 we can gain 2 dB at $10^{-4}$.

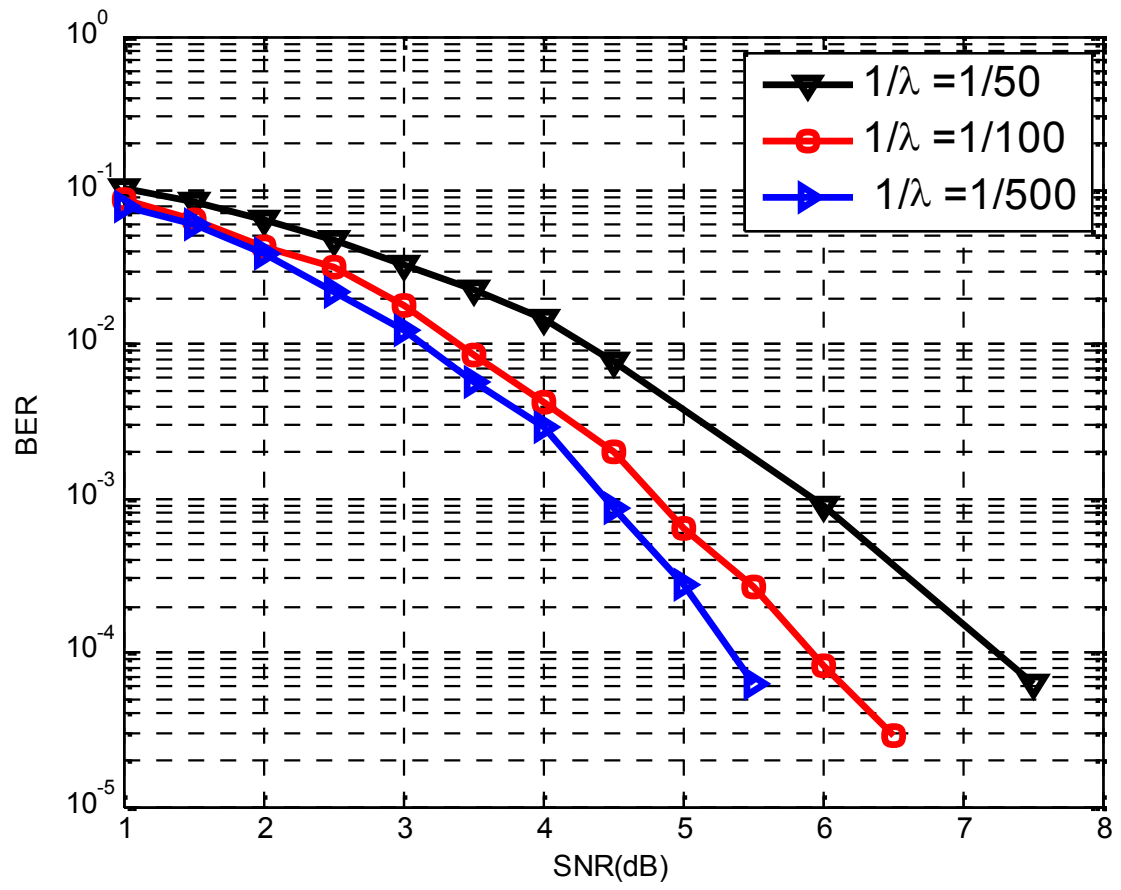

Fig. 3. Effect of step size on performances

### 3.2 Comparison of CGAD versus other decoding algorithms

In this subsection, we compare the performances of CGAD with other decoders. (Chase-2 decoding and Shakeel decoding algorithm).The performances of CGAD are better than Chase-2 algorithm as shown in Fig. 4 for BCH(127,113,5) code.

According to this figure, we observed that CGAD is comparable to Shakeel algorithm. CGAD outperforms chase-2 algorithm by 0.5 dB at $10^{-4}$ and is comparable to Shakeel algorithm as shown in Fig. 5.

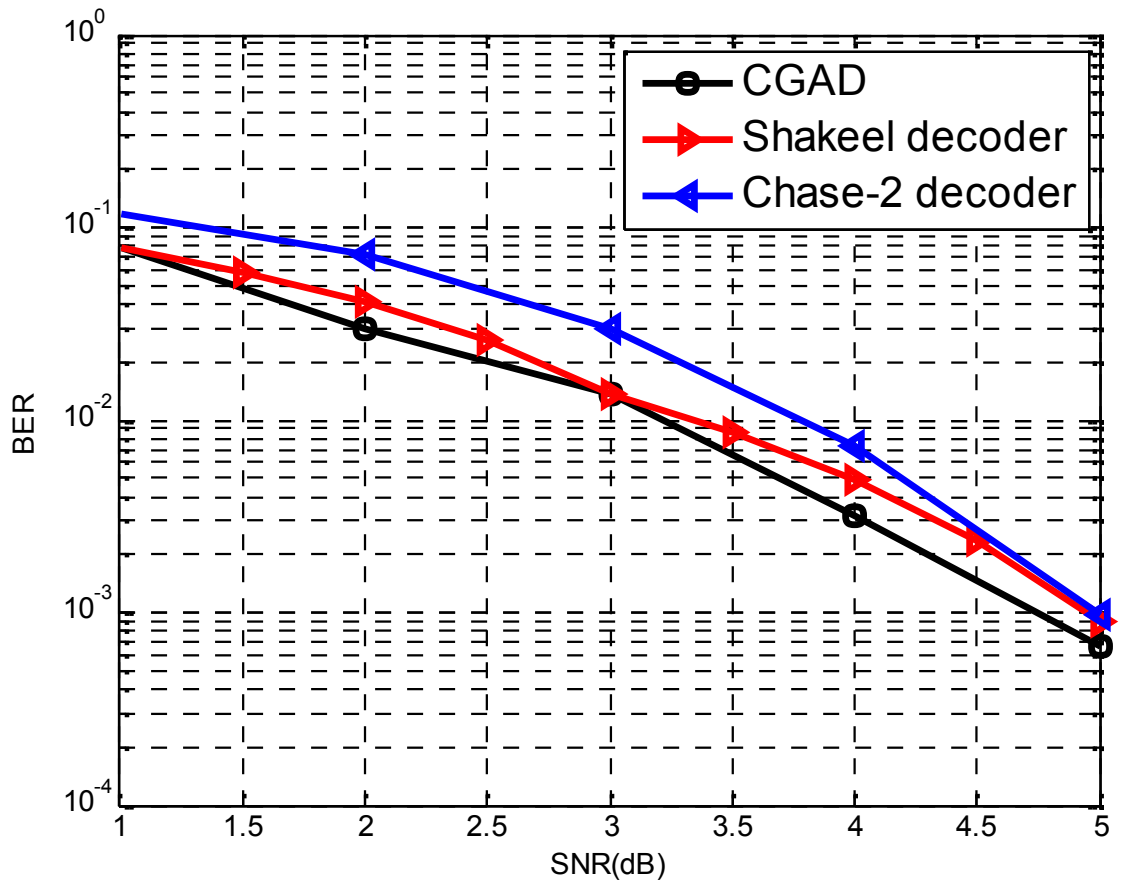

Fig. 6. Performances of Chase-2, CGAD and Shakeel algorithms for RS(15,7,9) code.

The Fig. 6 shows the performances of CGAD, Shakeel and Chase-2 algorithms for non-binary RS(15,7,9) code. We observe that our decoder is slightly better than the others algorithms at $10^{-3}$.

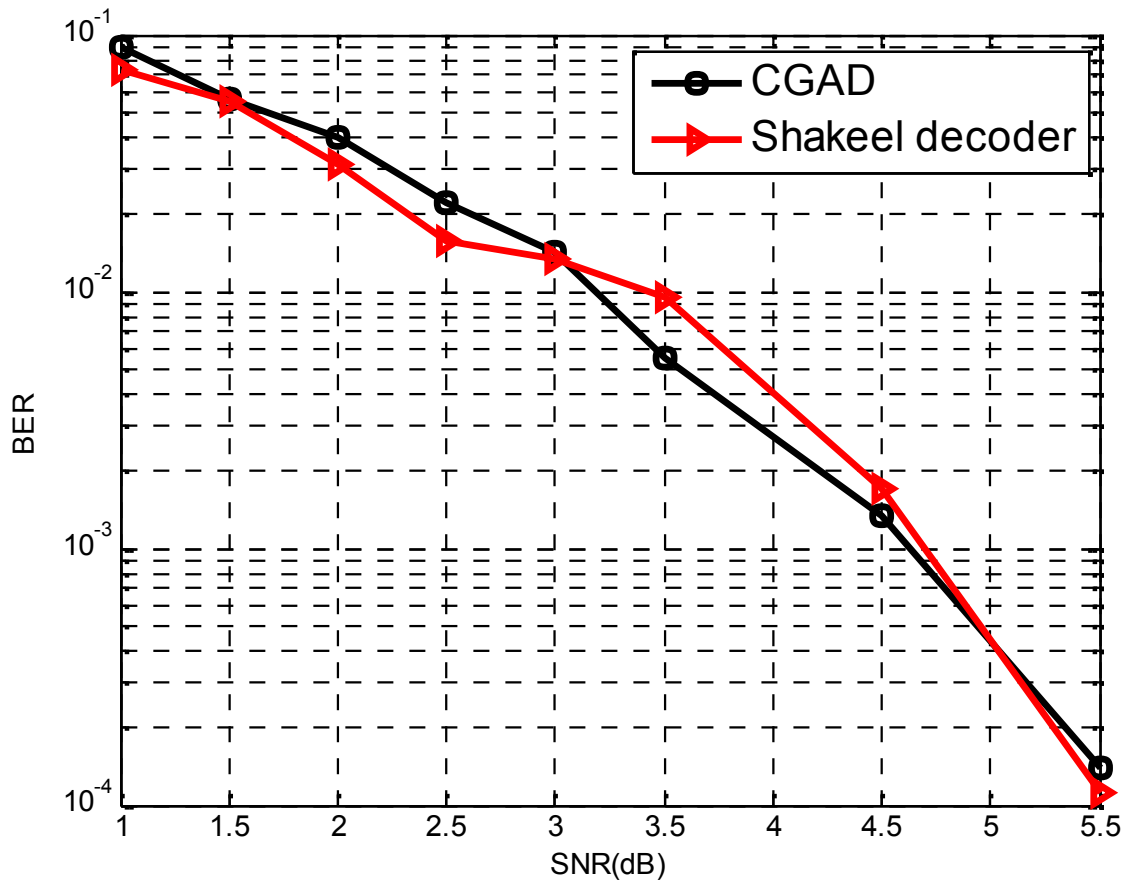

Fig. 7. Performances of CGAD and Shakeel algorithm for QR(71,36,11)

## 5. Complexity Analysis

### 5.1 Complexity Study

Let $n$ be the code length, $k$ the code dimension, and $t$ the error correction capability of a linear block code C. Let $T_c$ be the average number of generations. The Table 1 shows the complexity of the CGAD and Shakeel decoder. The complexity is polynomial in $k$, $(n-k)$ and $T_c$ for CGAD and $k$, $n$ and $T_c$ for Shakeel algorithm. We notice also that, our decoder is less complex than the Shakeel decoder for codes with a high rate $(n-k \prec\prec n)$.

Table 2: The complexity of CGAD and Shakeel algorithms

| Algorithm | Complexity |
|---|---|
| CGAD | $O(T_c k(n-k))$ |
| Shakeel decoder | $O(T_c kn)$ |

## 5.2 Experimental Study

### 5.2.1 Average number of generations required for convergence

The Fig. 8 shows the evolution of generations average number required for convergence (ANG) of CGAD and Shakeel algorithm versus SNR. We notice that the ANG decreases with increasing of SNR for CGAD, however it is stable for Shakeel algorithm (SA).

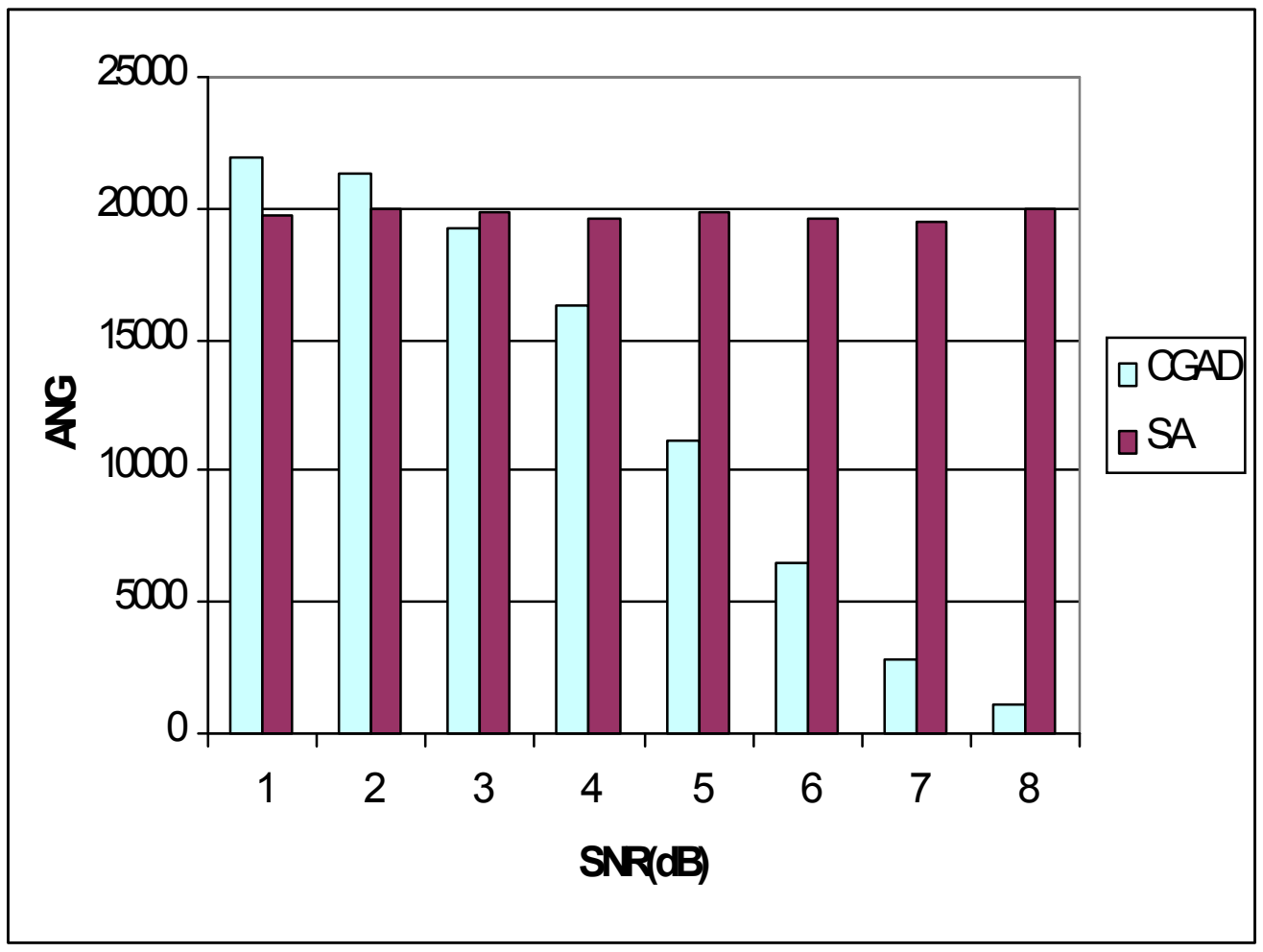

Fig. 8. The average number of generations required for convergence per SNR

### 5.2.2 Time Complexity

The Fig. 9 shows the time complexity of both genetic decoders. The time complexities are derived from calculating the run time (T) of 1000 blocks.
This figure shows that the time complexity of the Shakeel decoder is almost constant along the whole or SNR, for against it decreases with increasing SNR for our decoder. And it also shows that our decoder is less complex than the Shakeel algorithm (SA).

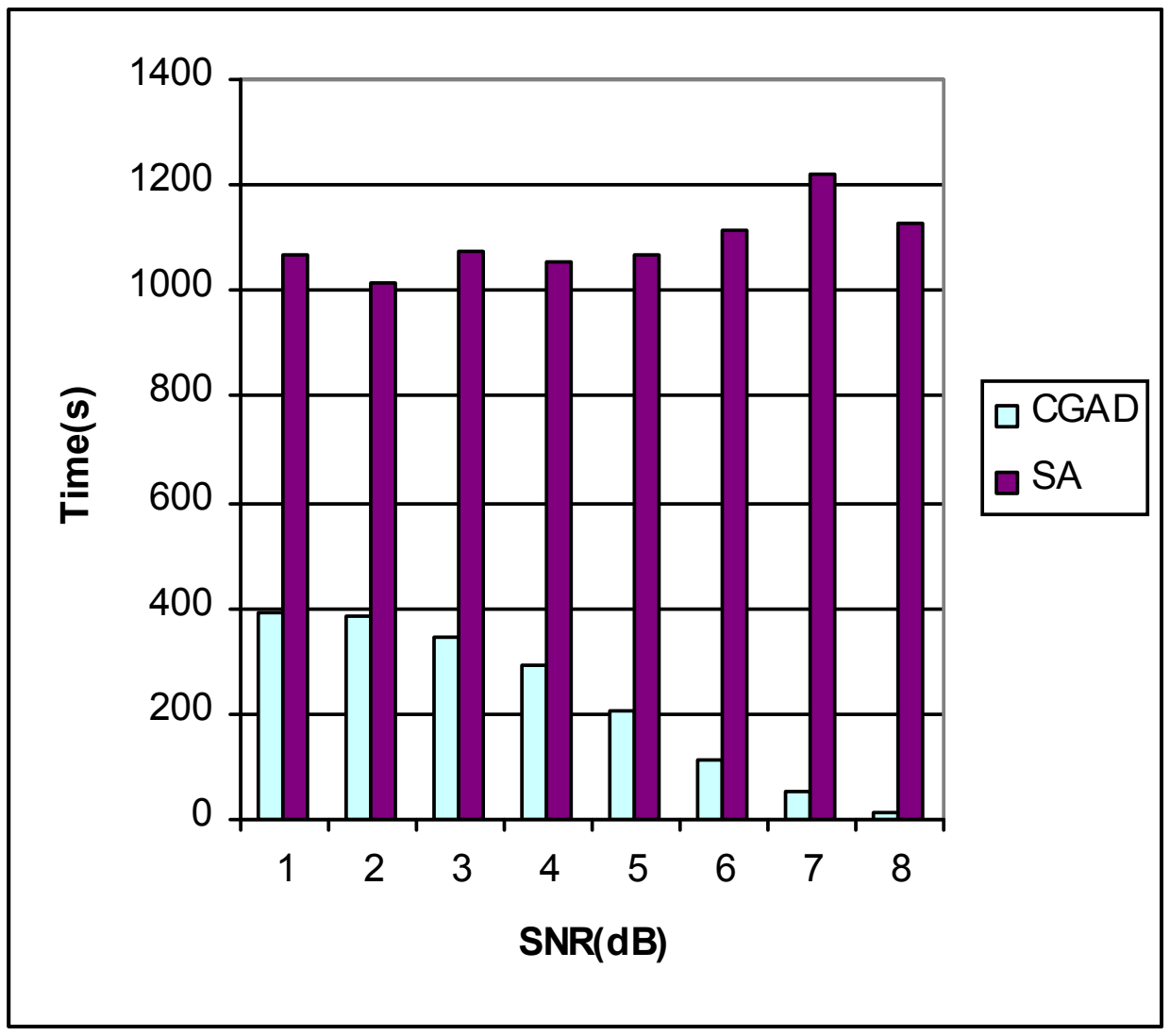

Fig. 9. The complexity of the two genetic decoders

## 5.3 Evolution of Hamming distance between two individuals with cGA generations.

The Fig. 10 illustrates the evolution of Hamming distance between the two individuals generated in step 4 of CGAD versus cGA generation's number (an average number for 1000 received block). It shows that the average Hamming distance is stable for the first 1000 generations, then it decreases until achieving zero value. This confirms the convergence of our algorithm.

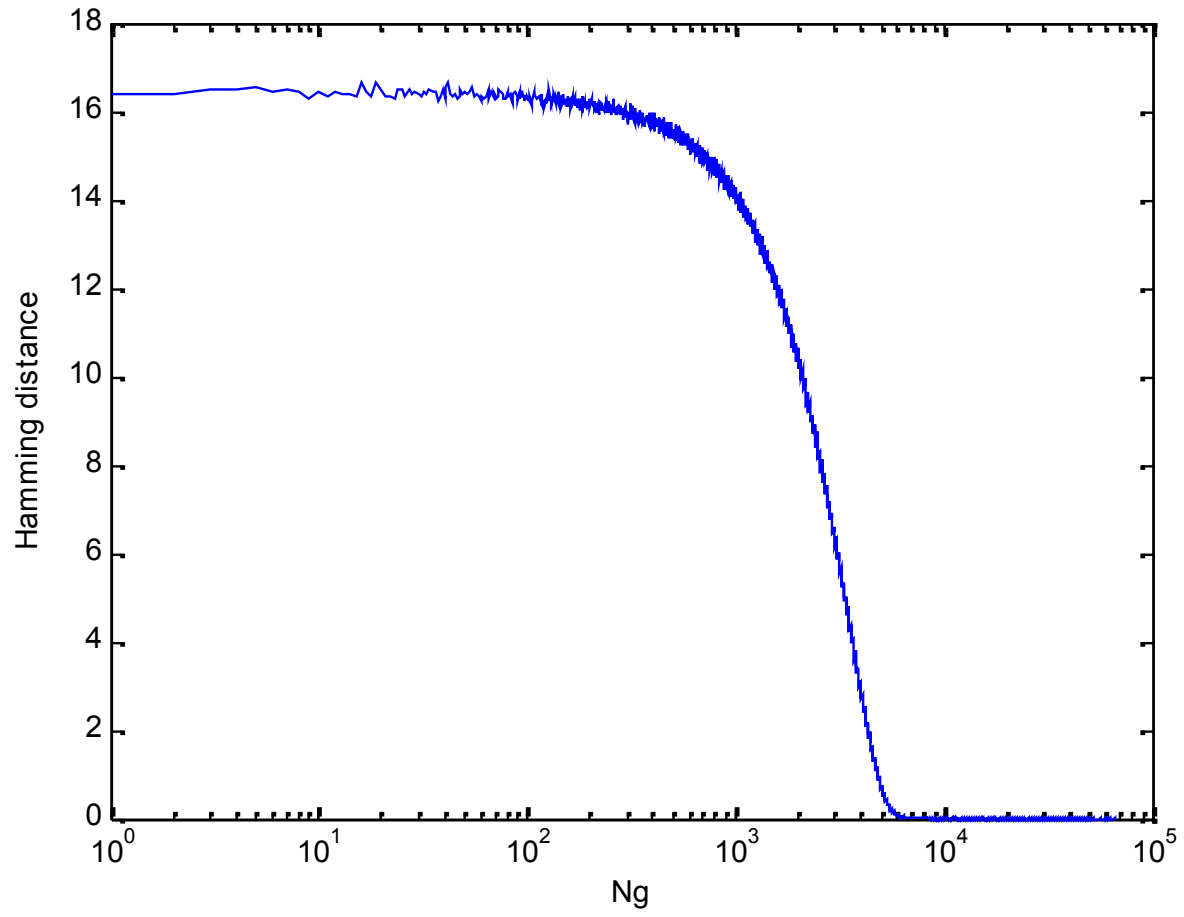

*Fig. 10 The Hamming distance between two individuals versus number of generations for CGAD.*

## 6. Optimized CGAD

According to our simulations, we observed that the CGDA takes more time to converge the last remaining positions of vector p, especially on the last position. Therefore, we thought to optimize our decoder by modifying the stopping criterion. The new algorithm (OCGAD) stops when there is a single position not yet converged (before the end of the convergence of any position vector). As a result, we assign the value 1 if the latter to its probability is greater than 0.5 and 0 otherwise.

### 6.1 Time complexity of OCGAD versus CGAD

The Fig. 11 shows the time complexity of CGAD and OCGAD. The time complexities are derived from calculating the run time (T) of 1000 blocks.
This figure shows that OCGAD presents a gain of about 40% in terms of time compared to CGAD.

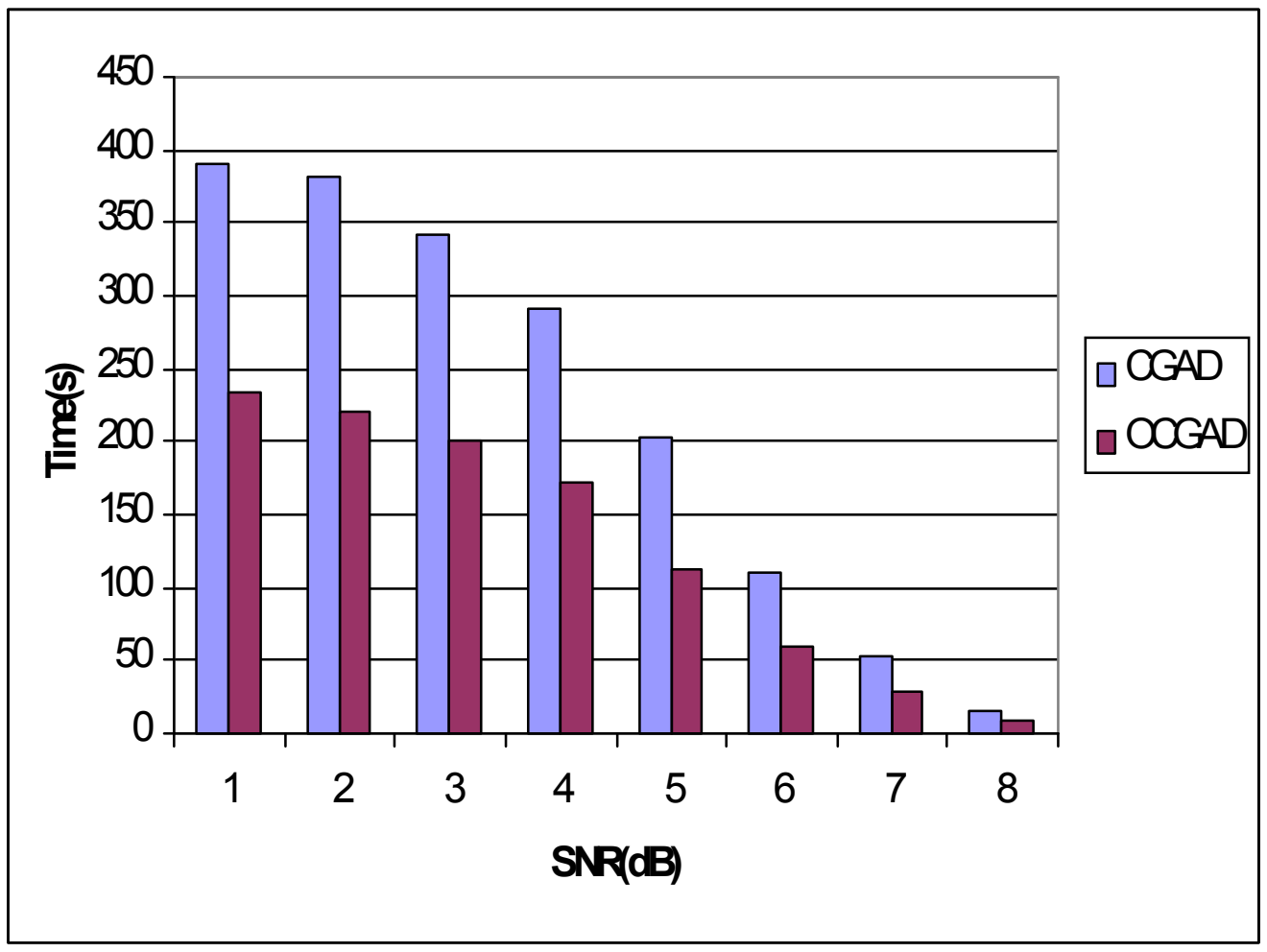

*Fig. 11. The time complexity of CGAD and OCGAD*

### 6.2 The average number of generations required for convergence

The Fig. 12 shows the evolution of generations average number required for convergence (ANG) of CGAD and OCGAD versus SNR. We notice that the ANG of OCGAD is smaller than ANG of CGAD for all SNR. (Reduction of approximately 40% of generations).

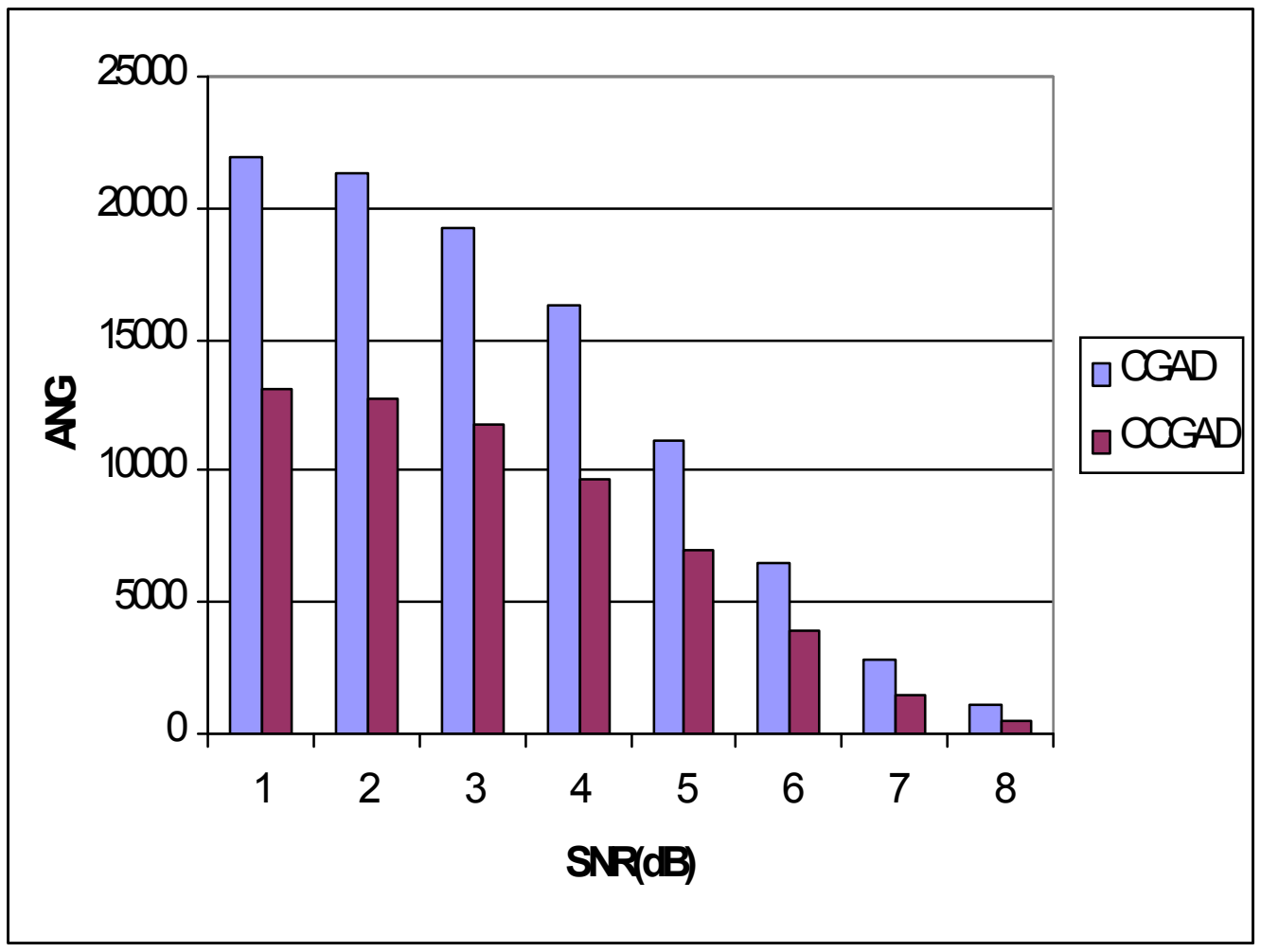

*Fig. 12 The average number of generations required for convergence per SNR of CGAD and OCGAD*

### 6.3 The performances of CGAD and OCGAD for BCH (63, 51, 5) code.

The Fig. 13 shows the performances of OCGAD and CGAD for BCH(63,51,5) code. We observe that OCGAD is slightly better than CGAD for SNR=6 dB.

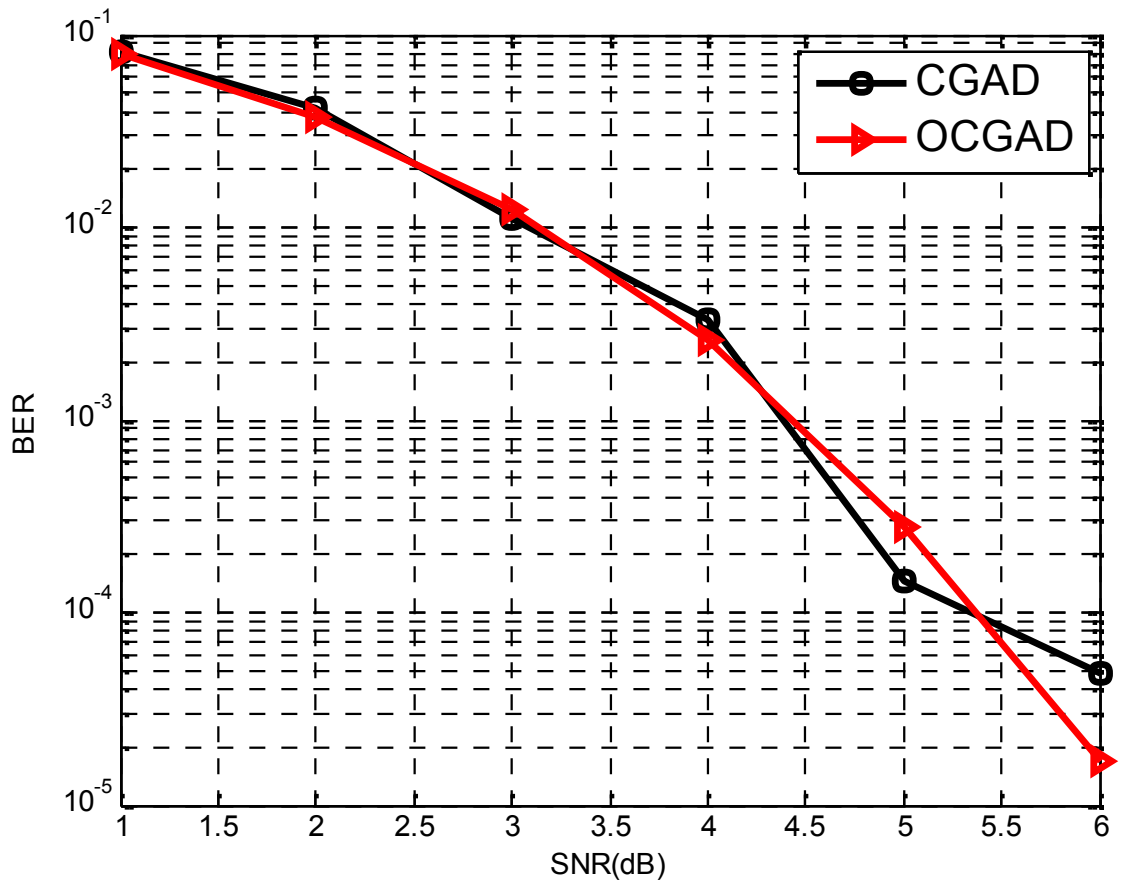

Fig. 13. The performances of CGAD and OCGAD for BCH(63,51,5) code

### 6.4 The performances of OCGAD and MLD for BCH(63,51,5) code.

The performances of OCGAD and MLD (Maximum Likelihood Decoding) for BCH(63,51,5) code, are shown in Fig. 14. From the simulation results, we observe that the performances of OCGAD are near that those of MLD.

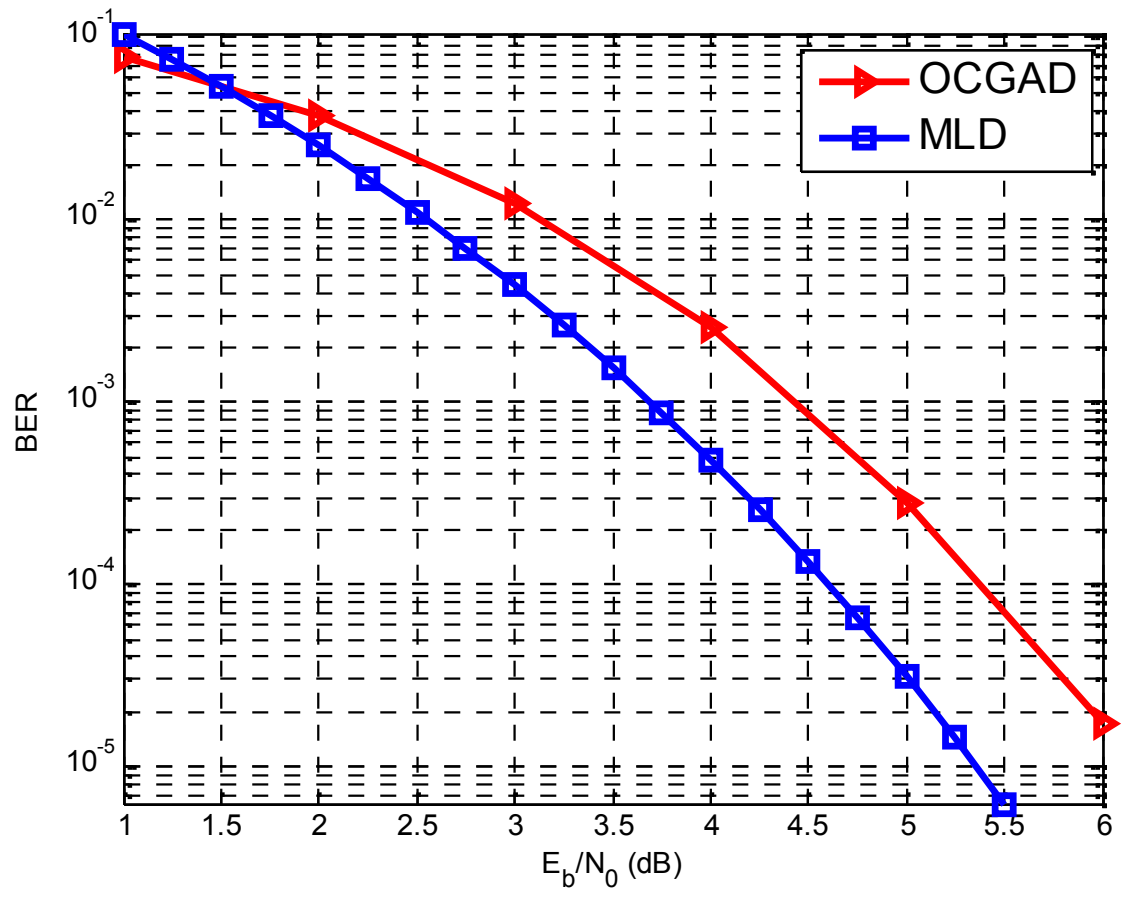

Fig. 14. The performances of CGAD and OCGAD for BCH (63,51,5) code

## 7. Conclusions

In this paper, we have proposed a new decoder based on cGA for linear block codes.

The simulations applied on some BCH, QR, and RS codes, show that the proposed algorithm is an efficient soft-decision decoding algorithm. Emphasis was made on the effect of the step size cGA parameter on the decoder performances. A comparison was done in terms of bit error rate performances and complexity aspects of the decoder. The proposed algorithm has an advantage compared to competitor decoder developed by Shakeel. Also, the optimized version of our algorithm has given good results in terms of performances and complexity.

The obtained results will open new horizons for the artificial intelligence algorithms in the coding theory field.

**Ahmed Azouaoui** received his license in Computer Science and Engineering in June-2001 and Master in Computer Science and telecommunication from University of Mohammed V - Agdal, Rabat, Morocco in 2003. Currently he is doing his PhD in Computer Science and Engineering at Department of Computer Science ENSIAS (Ecole Nationale Superieure d'Informatique et d'Analyse des Systems), Rabat, Morocco. His areas of interest are Information, Coding Theory and Artificial Intelligence.

**Ahlam Berkani** received his ingeneer diploma in Telecommunications and networks from ENSAO (Ecole Nationale des Sciences Appliquees d'Oujda), Morocco in 2010. Actually, she is preparing his PhD in Computer Science and Engineering at Department of Computer Science ENSIAS (Ecole Nationale Suprieure d'Informatique et d'Analyse des Systmes), Rabat, Morocco. His areas of interest are Information and Coding theory, and Artificial Intelligence.

**Pr. Mostafa Belkasmi** is a professor at ENSIAS (Ecole Nationale Superieure d'Informatique et d'Analyse des Systmes,Rabat); head of Telecom and Embedded Systems Team at SIME Lab.He had PhD at Toulouse University in 1991(France). His current research interests include mobile and wireless communications, interconnections for 3G and 4G, and Information and Coding Theory.